\documentclass[
    ,final            
  ]
  {aipproc}

\layoutstyle{6x9}
\usepackage{amsmath,amssymb,bm}

\begin{document}

\begin{flushright}
OU-HET-727
\end{flushright}

\title{Orbital angular momenta of quarks and gluons in the nucleon \ \    
-- model-dependent versus model-independent extractions --}

\classification{12.38.-t, 12.20.-m, 14.20.Dh, 03.50.De}
\keywords      {nucleon spin decomposition, gauge-invariance, evolution of gluon spin}

\author{Masashi Wakamatsu}{
  address={Department of Physics, Faculty of Science, Osaka University,
 Toyonaka, Osaka 560-0043, JAPAN}
}



\begin{abstract}
We demonstrate that there exist two kinds of gauge-invariant
decompositions of the nucleon spin, which are physically inequivalent.
The quark and gluon orbital angular momenta (OAMs) appearing in one
decomposition are basically the {\it canonical} OAMs, while the
quark and gluon OAMs appearing in another decomposition are called the
{\it dynamical} OAMs. It is shown that the dynamical OAMs of quarks and gluons
in the nucleon can be related to definite high-energy deep-inelastic-scattering
observables. On the other hand, we conjecture that the canonical OAMs of
quarks and gluon in the nucleon are model-dependent quantities, which is
meaningful only within a specific theoretical model of the nucleon.
\end{abstract}

\maketitle


\section{Introduction}

The current status and homework of the nucleon spin problem can
briefly be summarized as follows..
First, the quark polarization was fairly precisely determined
to be around $1/3$.
Second, gluon polarization is likely to be small, although with large
uncertainties. What carries the remaining 2/3 of the nucleon spin, then ?
Quark OAM ? Gluon spin ? Or gluon OAM ?  To answer this question
unambiguously, we cannot avoid to elucidate the following issues.
What is a precise definition of each term of the decomposition?
How can we extract individual term by means of direct measurements?
Especially controversy here are the OAMs of quarks and gluons.

\section{Model-independent extraction of quark and gluon orbital angular
momenta in the nucleon}

As is widely known, there have been two popular decompositions of the
nucleon spin. One is the Jaffe-Manohar decomposition \cite{JM90},
while the other is the Ji decomposition \cite{Ji97}.
%
%
%
In these popular decompositions, only the intrinsic quark spin part
is common, and the other parts are totally different.
An apparent disadvantage of the Jaffe-Manohar decomposition is that
each term is not separately gauge-invariant except for the
quark spin part. On the other hand, each term of the Ji decomposition is
separately gauge-invariant. Unfortunately, further gauge-invariant decomposition
of $J^g$ into its spin and orbital parts is given up in this well-known
decomposition.
An especially important observation here is that, since the quark OAMs
in the two decompositions are apparently different, 
one must necessarily conclude
that the sum of the gluon spin and OAM in the Jaffe-Manohar decomposition
does not coincide with the gluon total angular momentum in the Ji decomposition.


Some years ago, a new gauge-invariant decomposition of nucleon spin
was proposed by Chen et al. \cite{Chen08},\cite{Chen09}.
The basic idea is to decompose the gluon field $\bm{A}$ into
two parts, the physical part $\bm{A}_{phys}$ and
the pure-gauge part $\bm{A}_{pure}$.
Imposing some additional conditions, i.e. what-they-call the generalized
Coulomb gauge condition, Chen et al. arrived at the decomposition of
the nucleon spin in the following form : 
\begin{eqnarray}
 \bm{J}_{QCD} &=& \int \,\psi^\dagger \,
 \frac{1}{2} \,\bm{\Sigma} \,\psi \,\,d^3 x \ + \ 
 \int \psi^\dagger \,\bm{x} \times 
 ( \bm{p} - g \,\bm{A}_{pure} )
 \,\psi \,\,d^3 x \nonumber \\
 &+& \int \,\bm{E}^a \times 
 \bm{A}^a_{phys} \,d^3 x
 \ + \
 \int \,E^{aj} \,(\bm{x} \times \nabla ) \,
 \bm{A}^{aj}_{phys}  
 \,\,d^3 x \nonumber \\
 &=& \bm{S}^{\prime q} \ + \ 
 \bm{L}^{\prime q} \ + \ 
 \bm{S}^{\prime g} \ + \ 
 \bm{L}^{\prime g} .
\end{eqnarray}
An interesting feature of this decomposition is that each term is
separately gauge-invariant, while allowing the decomposition of the
gluon total angular momentum into its spin and orbital parts.
Chen et al's claim arose quite a controversy concerning the
definition of quark and gluon OAMs 
\cite{Chen08}\nocite{Chen09}\nocite{Cho10}\nocite{Chen11}\nocite{Leader11}
\nocite{Hatta11}\nocite{Wakamatsu10}\nocite{Wakamatsu11A}-\cite{Wakamatsu11B}.
In our opinion,
the only way to settle the controversy is to clarify a concrete
relationship  between the proposed decompositions and direct
observables. We believe that we have succeeded to reach this goal,
step by step, through the recent three papers
\cite{Wakamatsu10},\cite{Wakamatsu11A},\cite{Wakamatsu11B}.

In the 1st paper \cite{Wakamatsu10}, we have shown that the way
of gauge-invariant decomposition of nucleon spin is not necessarily unique,
and proposed yet another gauge-invariant decomposition given in the
following form :
\begin{eqnarray}
 \bm{J}_{QCD} &=& 
 \bm{S}^{q} \ + \ 
 \bm{L}^{q} \ + \ 
 \bm{S}^{g} \ + \ 
 \bm{L}^{g} ,
\end{eqnarray}
where
\begin{eqnarray}
 \bm{S}^q &=& \int \,\psi^\dagger \,
 \frac{1}{2} \,\bm{\Sigma} \,\psi \,\,d^3 x , \\
 \bm{L}^q &=& \int \,\psi \,
 \bm{x} \times
 (\bm{p} - g \,\bm{A} )
 \,\psi \,\,d^3 x , \\
 \bm{S}^g &=& \int \,
 \bm{E}^a \times 
 \bm{A}^a_{phys} \,\,d^3 x , \\
 \bm{L}^g &=& \int \,E^{aj} \,
 (\bm{x} \times \nabla) \,
 A^{aj}_{phys} \,\,d^3 x \ + \ 
 \int \,\rho^a \,(\bm{x} \times 
 \bm{A}^a_{phys} ) \,d^3 x . \label{Eq-Lg}
\end{eqnarray}
The characteristic features of our decomposition are as follows.
First, the quark parts of this decomposition is common with the Ji
decomposition. Second, the quark and gluon spin parts are common
with the Chen decomposition. A crucial difference with the Chen
decomposition appears in the orbital parts.
That is, although the sums of the quark and gluon OAMs in the two
decompositions are the same, i.e.
\begin{equation}
 \bm{L}^q \ + \ \bm{L}^g
 \ = \  
 \bm{L}^{\prime q} \ + \ 
 \bm{L}^{\prime g} ,
\end{equation}
each term is different in such a way that 
\begin{equation}
 \bm{L}^g - \bm{L}^{\prime g}
 \ = \ 
  - \,
 (\,\bm{L}^q - \bm{L}^{\prime q})
 \ = \ 
 \int \,\rho^a \,(\bm{x} \times 
 \bm{A}^a_{phys}) \,\,d^3 x .
\end{equation}
The difference arises from the treatment of the 2nd term of Eq.(\ref{Eq-Lg}),
which is {\it solely} gauge-invariant.
We call this term the {\it potential angular momentum} term, since
the QED correspondent of this term is the OAM
carried by the electromagnetic field (or potential), which appears
in the famous Feynman paradox raised in his textbook of classical
electrodynamics.
We have included this term into the {\it gluon} OAM part,
while Chen et al. included it into the {\it quark} OAM part.

In the 2nd paper \cite{Wakamatsu11A}, we made covariant extenstion of
gauge-invariant decompositions of the nucleon spin. 
Covariant generalization of the decomposition has twofold advantages.
First, it is essential to prove frame-independence of the decomposition.
Second, it generalizes and unifies the nucleon spin decompositions in the market. 
Basically, we find two physically different decompostions.
The decomposition (I) contains the well-known Ji decomposition, although it
also allows gauge-invariant decomposition of gluon total angular momentum
into its spin and OAM parts.
The decomposition (II) contains three known decomposition, i.e. those of 
Bashinsky-Jaffe \cite{BJ99}, of Chen et al. \cite{Chen08},\cite{Chen09},
and of Jaffe-Manohar \cite{JM90}, as will be shown below.

The startingpoint of our general analysis is a decomposition of the full
gauge field into its physical and pure-gauge parts, simiar to Chen et al.
Here, we impose only the following general conditions only.
The first is the pure-gauge condition for $A^\mu_{pure}$ : 
\begin{eqnarray}
 F^{\mu \nu}_{pure} \ \equiv \ 
 \partial^\mu \,A^\nu_{pure} - 
 \partial^\nu \,A^\mu_{pure} - i \,g \,
 [A^\mu_{pure}, A^\nu_{pure}] 
 \ = \ 0 ,
\end{eqnarray}
while the second is the
gauge transformation properties for these two components :
\begin{eqnarray}
 A^\mu_{phys}(x) &\rightarrow&
 U(x) \,A^\mu_{phys}(x) \,U^{-1}(x) , \\
 A^\mu_{pure}(x) &\rightarrow&
 U(x) \,\left(\,A^\mu_{pure}(x) - \frac{i}{g}
 \,\, \partial^\mu \,\right) \,U^{-1}(x) .
\end{eqnarray}
Actually, these condition alone are not enough to fix gauge
uniquely. However, the point of our analysis is that we can {\it postpone}
a complete gauge fixing until later stage, while accomplishing a
gauge-invariant decomposition of $M^{\mu \nu \lambda}$ based on the
above conditions only.

We start with the decomposition (II) given as
\begin{eqnarray}
 M^{\mu \nu \lambda}_{QCD} &=& 
 M^{\prime \mu \nu \lambda}_{q-spin} \ + \ 
 M^{\prime \mu \nu \lambda}_{q-OAM} \ + \ 
 M^{\prime \mu \nu \lambda}_{g-spin} \ + \ 
 M^{\prime \mu \nu \lambda}_{g-OAM} \nonumber \\
 &+& \ \mbox{\rm boost} 
 \ + \ 
 \mbox{\rm total divergence} ,
\end{eqnarray}
with
\begin{eqnarray}
 M^{\prime \mu \nu \lambda}_{q-spin}
 &=& \frac{1}{2} \,\epsilon^{\mu \nu \lambda \sigma} \,
 \bar{\psi} \,\gamma_{\sigma} \,\gamma_5 \,\psi , \\
 M^{\prime \mu \nu \lambda}_{q-OAM} 
 &=& \bar{\psi} \,\gamma^{\mu} \,(\,x^{\nu} \,i \,
 D^{\lambda}_{pure}  
 \ - \ x^{\lambda} \,i \,
 D^{\nu}_{pure} \,) \,\psi , \\
 M^{\prime \mu \nu \lambda}_{g-spin}
 &=& 2 \,\mbox{Tr} \,\{\, F^{\mu \lambda} \,A_{phys}^{\nu} \,
 \ - \ F^{\mu \nu} \,A_{phys}^{\lambda} \,\} , \\
 M^{\prime \mu \nu \lambda}_{q-OAM}
 &=& 2 \,\mbox{Tr} \,\{\, F^{\mu \alpha} \,
 (\,x^{\nu} \,D_{pure}^{\lambda}
 \ - \ x^{\lambda} \,D_{pure}^{\nu} \,) \,A_{\alpha}^{phys} \,\} .
\end{eqnarray}
At first sight, this decomposition appears a covariant
generalization of Chen et al's decomposition. However, a crucial difference is
that we have not yet fixed gauge explicitly. Owing to this general nature,
our decomposition (II) reduces to any ones of Bashinsky-Jaffe \cite{BJ99},
of Chen et al. \cite{Chen08},\cite{Chen09}, and of Jaffe-Manohar \cite{JM90},
after an appropriate gauge-fixing in a suitable Lorentz frame, which dictates that
these 3 decompositions are all {\it gauge-equivalent} !
They are not recommendable decompositions, however, because the quark
and gluon OAMs in those do not correspond to {\it known} experimental observables. 

Our recommendable is the decomposition (I) given as
\begin{eqnarray}
 M^{\mu \nu \lambda} &=& M^{\mu \nu \lambda}_{q - spin}
 \ + \ M^{\mu \nu \lambda}_{q - OAM} \ + \ M^{\mu \nu \lambda}_{g - spin}
 \ + \ M^{\mu \nu \lambda}_{g - OAM} \nonumber \\
 &+& \ \mbox{\rm boost}  
 \ \ + \ \ \mbox{total divergence} ,
\end{eqnarray}
with
\begin{eqnarray}
 M^{\mu \nu \lambda}_{q - spin}
 &=& M^{\prime \mu \nu \lambda}_{q-spin} , \\
 M^{\mu \nu \lambda}_{q - OAM}
 &=& \bar{\psi} \,\gamma^\mu \,(\,x^{\nu} \,i \,
 D^{\lambda} 
 \ - \ x^{\lambda} \,i \,
 D^{\nu} \,) \,\psi 
 \ \ \neq \ \  
 M^{\prime \mu \nu \lambda}_{q-OAM} , \\
 M^{\mu \nu \lambda}_{g - spin} 
 &=& M^{\prime \mu \nu \lambda}_{g-spin} , \\
 M^{\mu \nu \lambda}_{g - OAM} 
 &=& M^{\prime \mu \nu \lambda}_{g-OAM}
 \ + \ 2 \,\mbox{Tr} \,[\, (\,D_{\alpha} \,F^{\alpha \mu} \,)
 \,(\,x^{\nu} \,A^{\lambda}_{phys} \ - \ 
 x^{\lambda} \,A^{\nu}_{phys} \,) \,] .
\end{eqnarray}
It differs from the decomposition (II) in the orbitals parts.
The quark OAM part contains full covariant derivative contrary to the
decomposition (II). Correspondingly, the gluon OAM part is also different.
It contains a covariant generalization of the potential angular momentum term.
A superiority of this decomposition is that the quark
and gluon OAMs in this decomposition can be related to experimental observables ! 

Before demonstrating the superiority of the decomposition (I), it may be instructive
to confirm the fact that the difference between the quark and gluon OAMs
in the two decompositions is {\it nothing spurious}, i.e. it is {\it physical}.
This can, for instance, be convinced from an explicit model analysis
by Burkardt and BC \cite{BB09}. Using simple toy
models, i.e. scalar diquark model and QED and QCD to order $\alpha$, they
compared the fermion OAM obtained from the Jaffe-Manohar
decomposition \cite{JM90} and from the Ji decomposition \cite{Ji97}.
In our terminology, these two fermion OAMs are nothing but the {\it canonical} OAM 
and the {\it dynamical} OAM. Their findings are as follows. The two
decompositions give the same fermion OAMs in scalar diquark model, but
they do not in QED (gauge theory).
The $x$-distributions of fermion OAMs are different even in scalar diquark model.
In QED and QCD at order $\alpha$, two kinds of OAMs are definitely different.
Although their analysis is highly model-dependent, an important 
lesson to learn is that one should clearly distinguish two
kinds of OAMs, i.e. the canonical OAM (including its nontrivial gauge-invariant
extension due to Chen et al.) and the dynamical OAM, the difference of which
is nothing spurious, i.e. physical.

Now we shall explain why we recommend the decomposition (I). 
The keys are the following identities, which holds for the quark and gluon
OAM operators of our decomposition (I). For the quark part, we have
\begin{eqnarray}
 L_q &=& J_q \ - \ \frac{1}{2} \,\Delta q \nonumber \\
 &=& \frac{1}{2} \,\int_{-1}^1 \,x \,
 [\,H^q (x,0,0) \ + E^q (x,0,0) \,] \,dx \ - \ 
 \frac{1}{2} \,\int_{-1}^1 \,\Delta q(x) \,dx \nonumber \\
 &=& \langle p \uparrow \,| \,
 M^{012}_{q-OAM} \,| \,
 p \uparrow \rangle ,
\end{eqnarray}
with
\begin{eqnarray}
 M^{012}_{q-OAM} 
 \ = \ \bar{\psi} \,\left(\mbox{\boldmath $x$} \times \frac{1}{i} \,
 \mbox{\boldmath $D$} \right)^3 \,\psi
 &\neq& \left\{ \,
 \begin{array}{l}
 \bar{\psi} \,\left(\mbox{\boldmath $x$} \times \frac{1}{i} \,
 \nabla \right)^3
 \,\psi \\ 
 \bar{\psi} \,\left(\mbox{\boldmath $x$} \times \frac{1}{i} \,
 \mbox{\boldmath $D$}_{pure} \right)^3 \,\psi .
 \end{array}
 \right.
\end{eqnarray}
We find that the proton matrix element of our quark OAM operator coincides
with the difference between the 2nd moment of GPD $H + E$ and the 1st moment
of the longitudinally polarized distribution of quarks. What should be emphasized here
is that full covariant derivative appears, not a simple derivative operator nor its
nontrivial gauge-invariant extension.
In other words, the quark OAM extracted from the combined analysis of GPD and
polarized PDF is {\it dynamical} OAM (or {\it mechanical} OAM) not {\it canonical}
OAM. This fact is nothing different from Ji's claim. 

Similarly, for the gluon part, we find that the difference between the 2nd moment
of gluon GPD $H + E$ and the 1st moment of polarized gluon distribution coincides
with the proton matrix element of our gluon OAM operator given as follows : 
\begin{eqnarray}
 L_g &=& J_g \ - \ \Delta G \nonumber \\
 &=& \frac{1}{2} \,\int_{-1}^1 \,x \,[\,
 H^g (x,0,0) \ + \ E^g (x,0,0) \,] \,dx \ - \ 
 \int_{-1}^1 \,\Delta g(x) \,dx \nonumber \\
 &=& \langle p \uparrow \,| \,
 M^{012}_{g-OAM} \,| \,
 p \uparrow \rangle ,
\end{eqnarray}
with
\begin{eqnarray}
 M^{012}_{g-OAM} &=& 
 2 \,\mbox{\rm Tr} \,[\,E^j \,(\mbox{\boldmath $x$} \times 
 \mbox{\boldmath $D$}_{pure})^3 \,A^{phys}_j \,] 
 \hspace{10mm} : \ \ \ 
 \mbox{\rm canonical OAM}
 \nonumber \\
 &+& 2 \,\mbox{\rm Tr} \,[\,\rho \,(\mbox{\boldmath $x$} \times
 \mbox{\boldmath $A$}_{phys})^3 \,]
 \hspace{22mm} : \ \ \ 
 \mbox{\rm potential OAM term} .
\end{eqnarray}
Namely, the gluon OAM extracted from the combined analysis of GPD and polarized
PDF contains {\it potential} OAM term,  in addition to {\it canonical} OAM.
It would be legitimate to call the whole part the gluon {\it dynamical} OAM.

A natural next question is why the dynamical OAM can be observed ?  
An answer can be found in the famous textbook of quantum mechanics by
J.J. Sakurai \cite{BookSakurai95}.
There he discusses a motion of a charged particle in static electric
and magnetic field. It is emphasized that, under the electromagnetic
potential, the dynamical (or mechanical) momentum
$\bm{\Pi} \equiv m \,\frac{d \bm{x}}{dt}$, defined as a product of
mass and particle velocity, is given by $\bm{\Pi} \equiv \bm{p} - e \,\bm{A}$,
which is different from the canonical momentum $\bm{p}$.
Furthermore, what appears in the quantum version of Newton's equation of
motion is a dynamical momentum $\bm{\Pi}$ not a canonical one $\bm{p}$.
``Equivalence principle'' of Einstein dictates that the
{\it flow of mass} can in principle be detected by using gravitational force as a probe.
Naturally, the gravitational force is too weak to be used as a probe of
mass flow accompanying the motion of microscopic particle.
However, remember that the 2nd moments of unpolarized GPDs are also called the
gravito-electric and gravito-magnetic form factors. 
The fact that the dynamical OAM as well as dynamical linear momentum (fraction)
can be extracted from GPD analysis is therefore not a mere accident !

Our final comment is on the {\it quantum loop effects}, which have not been
considered in our analysis so far. Here, it may
be instructive to restart the whole argument with
a general reasoning deduced from the widely-accepted fact as follows.
By now, no one would object to the fact that
the nucleon spin can be decomposed into the total angular momenta of
quarks and gluons in a gauge-invariant way by means of GPD measurements.
Since the quark polarization is gauge-invariant and measurable through the
polarized DIS measurements, the quark OAM defined as a difference between the
total quark angular momentum and the longitudinal quark polarization is clearly
gauge-invariant and observable. The gluon part is a little more subtle.
If $\Delta G$ is really gauge-invariant and measurable, the gluon OAM defined
as a difference between the total gluon OAM and the gluon polarization should be gauge-invariant and observable.
Therefore, a key question is `Is $\Delta G$ really a gauge-invariant quantity
or not ?''

This is a very delicate question. In fact, it was often claimed that $\Delta G$
has its meaning only in the light-cone gauge and infinite-momentum frame.
More specifically, in an influential paper \cite{HJL99}, Hoodbhoy, Ji, and Lu
concluded that $\Delta G$ evolves differently in the Feynman gauge and the
LC gauge. 
However, the gluon spin operator used in their Feynman gauge calculation is
not gauge-invariant and delicately different from our gauge-invariant
gluon-spin operator.
The question is how to introduce this difference into the Feynman rule of
evaluating 1-loop anomalous dimension of the quark and gluon spin operators.
This problem was attacked and solved in our 3rd paper \cite{Wakamatsu11B}.
We find that the calculation in the Feynman gauge (as well as in any covariant
gauge including the Landau gauge) reproduces the answer obtained in the LC
gauge, which is also the answer obtained in the famous Altarelli-Parisi method. 

Our finding is important also from another context. 
So far, a direct check of the answer of Altarelli-Pasiri method for the evolution
of $\Delta G$ within the operator-product-expansion (OPE) framework was
limited to the LC gauge only, because it has been believed that there is no
gauge-invariant definition of gluon spin in the OPE framework.
This is the reason why the question of gauge-invariance of $\Delta G$ has
been left in unclear status for a long time !
Now we can definitely say that the gluon spin operator appearing
in our nucleon spin decomposition (although nonlocal) certainly provides us with
a completely satisfactory operator definition of gluon spin with full
gauge invariance, which has been searched for nearly 40 years. 

\section{Model-dependent insights into the quark and gluon orbital angular
momenta in the nucleon}

In the previous section, we have emphasized the existence of two kinds of
orbital angular momenta, i.e. the {\it canonical} and {\it dynamical} OAMs.
We argued that the {\it dynamical} OAM can be observed through the combined
analyses of unpolarized GPDs and longitudinally polarized PDFs. Is there any
possibility to extract {\it canonical} OAM by means of direct measurements ?
If this is possible, it means that we can isolate the very {\it correspondent}
of {\it potential angular momentum} appearing in the Feynman paradox.
Unfortunately, we are a little pessimistic about this possibility by the reason
explained below.

To explain it, we first recall a model-dependent sum rule for
the quark OAM in the nucleon advocated by Avakian et al. \cite{AESY0810}.
They showed that, within the framework of the MIT bag model (and also in
scalar diquark model), a certain weighted-integral of a T-even and chiral-odd
TMD distribution $h_{1T}^{\perp q} (x, \bm{k}_\perp^2)$,
called the pretzolocity, reduces to the OAM of quarks in the nucleon as
\begin{eqnarray}
 \langle L_3^q \rangle \ = \ - \, 
 \int \,dx \,\int \,d^2 \bm{k}_\perp \,\,\,
 \frac{\bm{k}_\perp^2}{2 \,M} \,\,\,
 h_{1T}^{\perp q} (x, \bm{k}_\perp^2) \label{Pretzolocity_SR} .
\end{eqnarray}   
%
Taking $Q = u + d$, we have
\begin{equation}
 \langle L^Q_3 \rangle \ = \ \frac{2}{3} \,\,P_P ,
\end{equation}
with
\begin{equation}
 P_S \ = \ \int_0^R \,[f(r)]^2 \,r^2 \,dr, \ \ \ 
 P_P \ = \ \int_0^R \,[g(r)]^2 \,r^2 \,dr .
\end{equation}
Here, $f(r)$ and $g(r)$ are the radial parts of the upper and lower components
of the ground state wave function of the MIT bag model.
The complete expression of the nucleon spin decomposition is also very simple
in the MIT bag model : 
\begin{eqnarray}
 \langle J_3 \rangle &=& 
 \langle L_3^Q \rangle
 \ + \ \frac{1}{2} \,\langle \Sigma_3^Q \rangle \ = \ 
 \frac{2}{3} \,\,P_P  \ + \ \frac{1}{2} \,
 \left(\, P_S - \frac{1}{3} \,\,P_P \,\right) \ = \ 
 \frac{1}{2}.
\end{eqnarray}
Unfortunately, MIT bag model is not a realistic model of the nucleon, which
is a bound state of nearly zero-mass quarks !
 Important physics like chiral symmetry is not properly
taken into account. More serious would be the neglect of gluon degrees of
freedom, which are widely believed  to carry sizable amount of nucleon
momentum fraction.
In any case, the above simple relation (\ref{Pretzolocity_SR}) obtained
in the MIT bag model must
be taken with care. The point is that the probability $P_P$ related to the
quark OAM is a highly model-dependent quantity.

A deeper meaning of our statement above may be understood by considering
much simpler composite system, i.e. the deuteron.
It is known that, in the simplest approximation, the magnetic moment of the
deuteron is given by the formula : 
\begin{eqnarray}
 \mu_d &=& 
 \mu_p \ + \ \mu_n \ - \ \frac{3}{2} \,\,
 P_D \,
 \left(\mu_p + \mu_n - \frac{1}{2} \right) .
\end{eqnarray}
Here, $\mu_p$ and $\mu_n$ are
the magnetic moments of the proton and neutron, while
$P_D$ is the so-called D-state probability of the deuteron.
The above formula indicates that, by measuring the magnetic moment,
one can extract the D-state probability of the deuteron.
This D-state probability also gives the measure of OAM contents
of the deuteron, as is clear from the following angular momentum
decomposition of the deuteron spin. 
\begin{eqnarray}
 \langle J_3 \rangle &=& 
 \langle L_3 \rangle 
 \ + \ \langle S_3 \rangle \ = \ 
 \frac{3}{2} \,\,P_D \ + \ 
 \left(\,P_S - \frac{1}{2} \,\,P_D \,\right)
 \ = \ 1 .
\end{eqnarray} 
However, it is a well-known fact that the D-state probability of the deuteron,
which is thought to be an analogous object to the probability $P_P$ of the
MIT bag model, is not a direct observable \cite{Amado79},\cite{Friar79}.
We emphasize that the OAM, which we discuss above corresponds to
an expectation value of {\it canonical} OAM operator
between some Fock-state eigenvectors, which has a definite meaning only
within a specific theoretical model. 

\section{Summary and conclusion}

We have discussed the OAM of composite particles, with particular emphasis upon
the existence of two kinds of OAM, i.e. canonical OAM and dynamical OAM as
well as canonical momentum and dynamical momentum.
The canonical momentum is certainly a fundamental building block in the theoretical
framework of quantum mechanics and quantum field theory. However, whether it
corresponds to an observable is a different thing !
In contrast, we have shown that the dynamical OAM of quarks and gluons
in the nucleon
can in principle be extracted model-independently from combined analysis of
the GPD measurements and the polarized DIS measurements.
This means that we now have a satisfactory theoretical basis toward a complete decomposition of the nucleon spin, which is a strongly-coupled relativistic bound
system of quarks and gluons.

\begin{theacknowledgments}
 The author is very grateful to A.W.~Thomas for his kind hospitality
during the PacSpin 2011 at Cairns, Australia.
\end{theacknowledgments}



\bibliographystyle{aipproc}   

\bibliography{sample}

\IfFileExists{\jobname.bbl}{}
 {\typeout{}
  \typeout{******************************************}
  \typeout{** Please run "bibtex \jobname" to optain}
  \typeout{** the bibliography and then re-run LaTeX}
  \typeout{** twice to fix the references!}
  \typeout{******************************************}
  \typeout{}
 }

\end{document}